\documentclass[12pt]{iopart}
\usepackage{epsf}
\usepackage{setstack}
\usepackage{graphicx}

\def\(({\left(}
\def\)){\right)}
\def\[[{\left[}
\def\]]{\right]}

\newcommand{\be}{\begin{equation}}
\newcommand{\ee}{\end{equation}}
\newcommand{\bea}{\begin{eqnarray}}
\newcommand{\eea}{\end{eqnarray}}

\begin{document}

\title[Aging, memory and rejuvenation in simple models] {Aging, memory
  and rejuvenation: some lessons from simple models}

\author{Florent Krzakala$^a$ and Federico Ricci-Tersenghi$^b$}

\address{$^a$Lab. P.C.T., UMR CNRS 7083, ESPCI, 10 rue Vauquelin,
Paris 75005, France} \address{$^b$Dip. di Fisica, Universit\`a di Roma
"La Sapienza", P.~A.~Moro 2, Roma 00185, Italy}

\begin{abstract}
  Many recent experiments probed the off equilibrium dynamics of spin
  glasses and other glassy systems through temperature cycling
  protocols, and observed memory and rejuvenation phenomena. Here we
  show through numerical simulations, using powerful algorithms, that
  such features can already be observed to some extent in simple
  models such as two dimensional ferromagnets. We critically discuss
  these results and review some aspects of the literature in the light
  of our findings.
\end{abstract}

\maketitle

\section{Introduction}
One of the main field of research in ill-condensed matter over the
last few years was certainly the off equilibrium dynamics of glassy
systems. These studies led to the emergence of satisfying pictures and
useful concepts that now allow a good qualitative understanding of
many experimental facts, such as aging~\cite{Review_JP,Aging_Review}.
For instance, if one quenches a glassy system from its high
temperature phase to its low temperature phase, this system will {\it
age} and the longer the experimentalist will wait, the slower the
system will be (which is indeed what aging is in real life).  To be
more precise, after a quench at $t=0$, an observable like the magnetic
susceptibility under a field applied at time $t_w$ ---the so-called
waiting time--- will typically decay following a scaling function
$f(t/t_w)$. This generic picture of aging is now well documented and
quite ubiquitous, being observed in many experimental and theoretical
situations~\cite{Review_JP,Aging_Review}. However, one of the most
striking feature in the dynamics of these systems, which is not well
taken into account so far, is certainly their dependence to the
complete history, so that more complex procedures than a simple quench
are of great interest.  Indeed, following the early seminal work of
Struick and Kovacs~\cite{Struick}, a number of more elaborate
experiments have been performed in a wide class of glassy materials
such as polymers~\cite{Polymers}, colloidal suspensions under a
shear~\cite{Colloids}, disordered or frustrated
magnets~\cite{ReEntrant,NoDisorder} or, for what will matter here,
spin glasses~\cite{shift,Review_cycle,Review_houches}. Interesting and
impressive hysteresis effects have been observed; they are commonly
referred to as {\it memory} and {\it rejuvenation}.

Let us briefly discuss these effects in the context of spin glasses
(and refer for instance to
\cite{Review_JP,shift,Review_cycle,Review_houches,Ghost} for a more
exhaustive description). In standard experiment, a temperature cycle
is performed: a system (with a glass transition at $T_g$) is at first
quenched from its high temperature phase to $T_1<T_g$ and then kept at
this temperature for a while, before being cooled again to $T_2<T_1$.
After another time interval, it is brought back to $T_1$.  Two
striking effects are observed. 1) As the system is brought to $T_2$
its dynamics witnesses a large restart, although it looked almost
equilibrated at $T_1$; in particular its susceptibility is initially
much larger than what would be after the same time in a direct quench
at $T_2$. In the aging phenomenology a system that responds more is
{\it younger} thus the name {\it rejuvenation} for this effect that
has been observed in many materials~\cite{Polymers,Colloids,shift}.
2) After the stage at $T_2$, when brought back to $T_1$, the system
may behave (depending on the material and/or the parameters of the
experiment) as if the temperature cycle has not been done at all and
its susceptibility seems just to follow the $T_1$ curve from where it
was left in the first stage at $T_1$. In AgMn spin glass for
instance~\cite{Ghost}, there are no differences (apart from a short
transient) between the susceptibility obtained for a long quench at
$T_1$, and the one obtained in a temperature cycle if one just removes
by hand all the data corresponding to the time spent at $T_2$.  This
is called the {\it memory} effect as the system, despite its
rejuvenation, remembers how it was when it left $T_1$.

It is fair to say that we are still far from a complete theoretical
understanding of these two effects and their co-existence, apart from
simple phenomenological
descriptions~\cite{Review_cycle,Ghost,Review_houches,Picco00,JP_Ludo,Maiorano}.
Numerical simulations would be of great help in the understanding of
this problem, but at the moment they have produced a number of
contradictory claims: while some authors~\cite{Ghost} advocate
also for off-equilibrium typical configurations the presence of a
property called temperature chaos~\cite{BMchaos} (equilibrium
configurations at different temperatures are completely reshuffled for
sufficiently large systems), others claim to observe these
effects~\cite{JP_Ludo,Jimenez} in the absence of any chaos, a
conclusion that has also been challenged~\cite{Maiorano}. It was
even provocatively asked if the Edwards-Anderson spin glass model was
able to reproduce experimental findings~\cite{Picco00}, or if other
models would be more appropriate~\cite{XY_Ludo}. Many questions were
raised by the interpretation of numerical simulations, and in this
situation it is natural for a physicist to come back to what he knows
best: the ferromagnetic models we simulated in our early courses.
Doing such quenches and $T$-cycling simulations in the $2d$ Ising and
$XY$ models will indeed provide some interesting
lessons~\cite{FloFede} as we shall now discuss.

\section{Models and methods}
We consider $T$-cycle experiments ($T=\infty \rightarrow T_1<T_g \rightarrow
T_2<T_1 \rightarrow T_1$) in Monte Carlo (MC) simulations.  We use large
system sizes (typically $L\approx 10^3$) to avoid finite size effects and
equilibration.  We consider two models defined on a $2d$ square lattice: the
first has Ising spins and Hamiltonian $H = - \sum S_i S_j$, and the second has
2-component vector spins of unit length and Hamiltonian $H=-\sum \vec{S_i}
\vec{S}_j$, where the sums act on neighboring spins.  While the Ising model
undergoes a standard second order ferromagnetic transition at $T_c$ , the XY
model possesses a remarkable quasi-ordering characterized by a line of
critical points going from $T=0$ up to a transition temperature $T_{KT}$
\cite{KT}.  Finally, we briefly discuss finite dimensional spin glasses with
Gaussian random couplings (where we use, for $4d$, $L \approx 20$). We express
all temperatures in units of $T_c$ or $T_{KT}$ and consider Glauber as well as
Kawasaki dynamics.
  
A few words on the numerical methods used in this work. So far
simulations computed magnetic susceptibilities from correlation
functions, assuming the validity of the Fluctuation-Dissipation
Theorem (FDT); we will see that this can be sometime quite dangerous
at short times, when the system is still strongly out of
equilibrium. Instead, we used a generalization of a recently proposed
algorithm to compute directly the linear response to a (DC or AC)
magnetic field without {\it physically} putting the
field~\cite{ChatelainFede,GKR}. We will unfortunately skip here these
quite technical, but important, points (addressing the reader to a
more detailed paper~\cite{FloFede}) and instead will focus on the
results.

\section{Effective temperatures in coolings and heatings}

Let us start by few remarks on coolings from a initial high
temperature $T_i$ to a final lower one $T_f$ and on heatings from a
low temperature $T_i$ to a higher one $T_f$.  We concentrate on the
$2d$ XY model (that will be useful later on) where some analytical
results can be obtained in both cases, when the system is initially
equilibrated at $T_i$ (under the so-called {\it spin waves
approximation}~\cite{XY_dyn}).  One can show that the correlation
between a configuration at times $t_w$ and $t>t_w$
scales~\cite{XY_dyn} as
\be\label{MyResult}
C(t,t_w) \propto \((1 / t_w \))^{\frac{\eta(T_f)}{2}} \(( 1+
\frac{1}{4 \frac{t}{t_w}(\frac{t}{t_w} +1)}
\))^{\frac{\eta(T_f)-\eta(T_i)}{4}}\;,
\ee
where $\eta(T)$ is a critical exponent, roughly proportional to the
temperature T~\cite{XY_dyn,Bray_Review}. The use of the FDT allow us to
estimate the magnetic susceptibility at time $t$ to a field applied at time
$t_w$ or, more conveniently, the susceptibility under an oscillatory field of
frequency $\omega\approx1/t_w$: roughly $\chi(\omega,t) \approx
(1-C(t+1/\omega,t))/T$.  Using expression (\ref{MyResult}) we see that
starting from a high temperature $T_i$ and cooling down to $T_f<T_i$, then
$\eta(T_f)<\eta(T_i)$ and therefore $C(t+1/\omega,t)$ increases with $t$. As a
consequence, when the system ages one observes that its susceptibility
decreases towards its equilibrium value, as it is well known. However, when
heating from a low temperature $T_i$ to $T_f>T_i$ one has now
$\eta(T_f)>\eta(T_i)$ and $C(t+1/\omega,t)$ then decreases with $t$ and thus
the susceptibility increases with $t$; in this case one observes a kind of
{\it inverse aging} where the system is initially too correlated for the new
temperature $T_f$, so that it has to uncorrelate with time.

At short times after a change of temperature from $T_i$ to $T_f$ the
system is strongly out of equilibrium and the FDT is violated, so that
an effective temperature $T_{eff}$~\cite{FDT_violation} can be
defined.  Computing it in the Langevin formalism we found~\cite{FloFede}
\be
T_{eff} = T_f \((1+ \frac{1}{\omega t} \frac{T_i-T_f}{T_f}\)) = 
T_f \((1 - \frac{1}{\omega t}\)) + T_i \frac{1}{\omega t}\;.
\ee
This shows that, although $T_{eff}=T_f$ at large time (where FDT is
valid), at shorter time, when $t = \mathcal{O}(1/\omega)$, $T_{eff}$
is a weighted average of $T_i$ and $T_f$.  All of that is in fact
completely general and we will see from our data that this scenario
holds equally well for ferromagnets and for spin glasses: the moral of
this story is that FDT overestimates the real susceptibility in
coolings, and underestimates it in heatings.

\section{Temperature Cycle experiment in Ising model}

\begin{figure}
$^{({\rm a})}$\hspace{-1cm}
\includegraphics[width=0.53\textwidth]{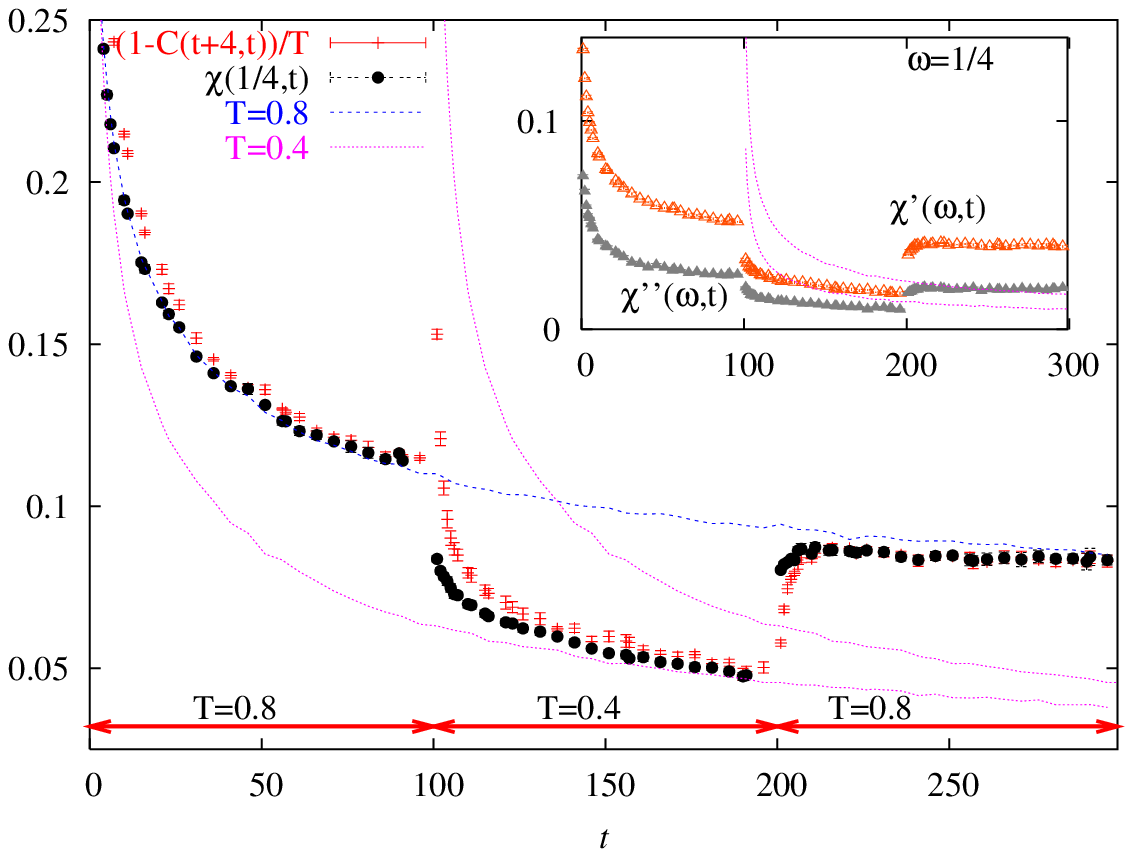}
\hspace{0.5cm}$^{({\rm b})}$\hspace{-1.1cm}
\includegraphics[width=0.53\textwidth]{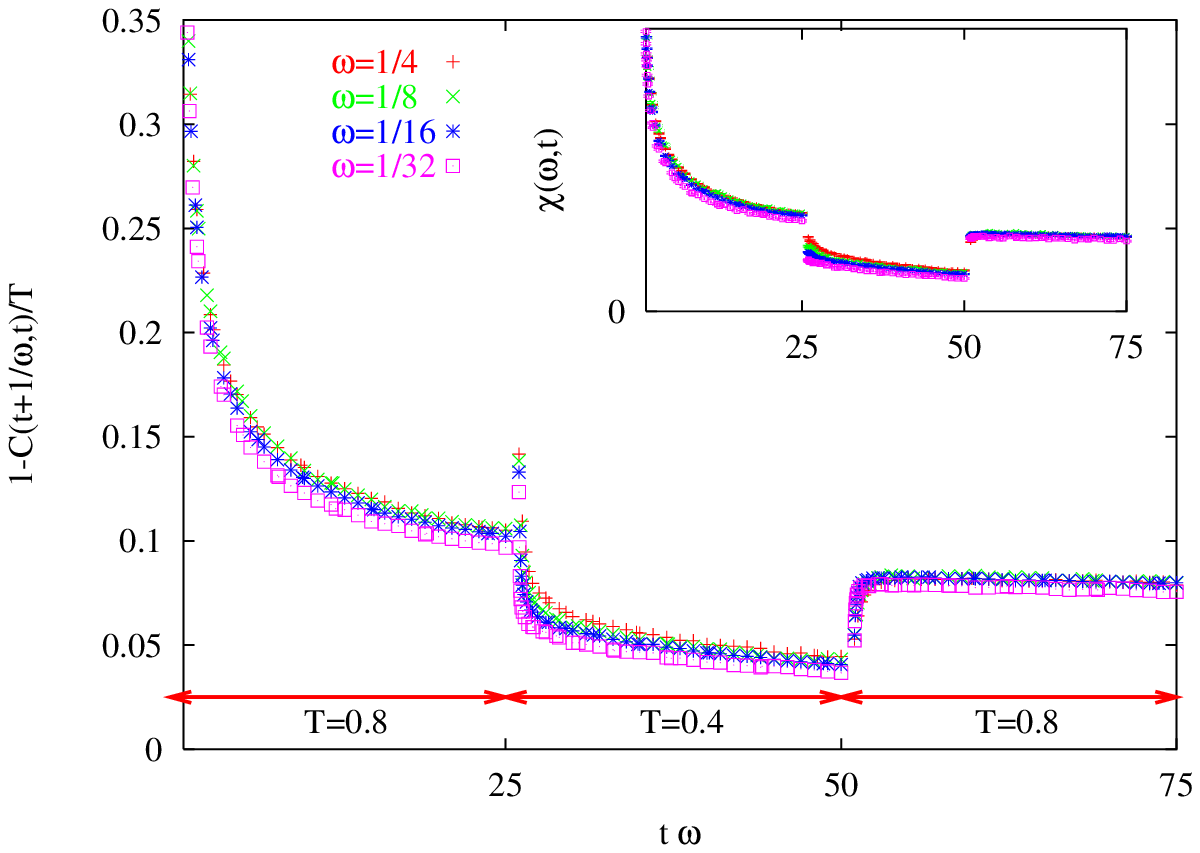}
\caption{\label{fig:1} $T$-cycling experiments in the $2d$ Ising model
  with Glauber dynamics.\\ (a) Susceptibility at time $t+1/\omega$
  under a DC field applied at time $t$, by assuming FDT, i.e.\
  $(1-C(t+1/\omega,t))/T$, and using the exact algorithm. The inset
  shows $\chi'$ and $\chi''$ obtained with the exact algorithm for AC
  field of frequency $\omega$.  A transient rejuvenation effect is
  observed upon cooling but no memory (full lines are data for direct
  quenches to $T_1$ and $T_2$).  (b) Same data for larger time scales,
  i.e.\ smaller $\omega$, as a function of the rescaled time $t
  \,\omega$, using FDT (main plot) and the exact algorithm (inset). As
  all times are rescaled rejuvenation vanishes fastly in the exact
  susceptibility while a transient signal is still present assuming
  FDT.}
\end{figure}

We now turn our attention to the numerical data in Fig.\ref{fig:1}(a),
obtained from a T-cycling simulation in the $2d$ Ising model with
Glauber dynamics. A clear restart of the aging dynamics (a
rejuvenation) is observed when cooling from $T_1=0.8$ to $T_2=0.4$,
while no memory effect is seen (going back to $T_1$, the
susceptibility is lower than the last point in the first stage at
$T_1$). Apart from this lack of memory (that we will discuss in the
next paragraph), this looks amazingly, and perhaps surprisingly,
similar to the curves obtained in cycling simulations of spin
glasses~\cite{JP_Ludo,Jimenez}; this rises the following questions.

First, should we be really surprised? After all, the temperature is changed so
that {\it something} has to happens as the system tries to equilibrate in the
new environment. Yet, it is easy to show in a simulation that this
equilibration will be very fast, and hardly observable, if we would start from
a state with spatially homogeneous magnetization.  However, we know that after
a quench the system is far from equilibrium. It is composed by domains of
positive and negative magnetizations separated by interfaces ---or domain
walls--- that grow with time~\cite{Bray_Review}. While the bulk part of
domains is indeed equilibrating almost instantaneously to the new temperature,
the interface itself needs more time to re-equilibrate and this is the origin
of the signal observed in Fig.\ref{fig:1}(a). The reader then may find that
this is a bit a trivial effect. How could this looks so impressive in
Fig.\ref{fig:1}(a)? We even observe that the susceptibility estimated by the
FDT is initially {\it larger} at $T_2$ than at $T_1$, and we would thus be
tempted to think that this is the sign of a very strong rejuvenation effect
(as it is sometimes claimed in the literature~\cite{Jimenez,Strong}). This is
not completely true.  Indeed, it is possible to show~\cite{FloFede} that at
large $\beta=1/T$, while the asymptotic equilibrium susceptibility
$\chi_{bulk}$ behaves as $\beta e^{-\beta}$, the aging part $\chi_{dw}$ due to
the domains walls scales as $\beta$, essentially because there are spins in
zero local field on these walls (this is in fact nothing else than the
division by $T$ in the dynamic FDT formula).  $\chi_{dw}$ is therefore very
sensitive to $T$-changes so that $\chi(T_2)$ can be made arbitrary high by
lowering $T_2$, while nothing really changes in the physics of the system. The
relative high of the susceptibility at different temperatures is therefore not
a very good measure for a restart of the dynamics. Finally, we can check that,
as predicted in the last section, we strongly overestimate ({\it resp.}
underestimate) the early time regime of the susceptibility upon cooling ({\it
  resp.}  heating) using the FDT, which therefore enhances artificially the
rejuvenation effect (a comment also made in \cite{Jimenez}).  The reason for
that can be easily understood: when a given spin is strongly out of
equilibrium (for instance when the temperature is changed), it will be forced
to flip and this will affect the correlation function. However, the
susceptibility is only sensitive to flips due to thermal fluctuations, not to
those driven by off-equilibrium relaxation: this is the origin of the large
discrepancy between the FDT approximate and the exact
susceptibility~\cite{FloFede}.

Still, a signal is observed in the exact susceptibility, so that the
puzzled reader may rightfully ask why then is there no rejuvenation in
{\it real} ferromagnets? First of all, this is not completely true as
rejuvenation is observed in some particular class of frustrated
magnets~\cite{ReEntrant}, but the answer to this question is that
before claiming any experimental relevance one has to do numerical
simulations on the same time scales than experiments, i.e.\ in the $t
\to\infty$ limit.  As can be seen in Fig.\ref{fig:1}(b), the
rejuvenation signal tends to vanish if one rescales all time scales by
the period $P$ of the oscillating field and send $P \to \infty$
(because the time needed to re-equilibrate interfaces is finite).
Again, notice that while this is clear on the correct susceptibility,
the approximate one still reports a misleading remaining signal in the
$P \to\infty$ limit.  Observing a plot such as the one in
Fig.\ref{fig:1}(a) is thus meaningless without a systematic large time
study.  These points obviously weaken many conclusions that have been
obtained from simulations so far.

\section{The magic of 2d Kawasaki dynamics}

While we observed a (transient) rejuvenation in Fig.\ref{fig:1}(a),
memory is lacking. This is easily understood: the coarsening dynamics
is almost $T$-independent, domains grows as $\sqrt{t}$ at any low
temperatures, and thus the susceptibility measured coming back at
$T_1$ is lower than the one the system had when it first left $T_1$:
essentially there are much less domain walls!  If, however, the
coarsening at $T_2<T_1$ is {\it much slower}, so that the density of
interface would not decreases too fast, we may expect a good memory
effect.

\begin{figure}
$^{({\rm a})}$\hspace{-1cm}
\includegraphics[width=0.53\textwidth]{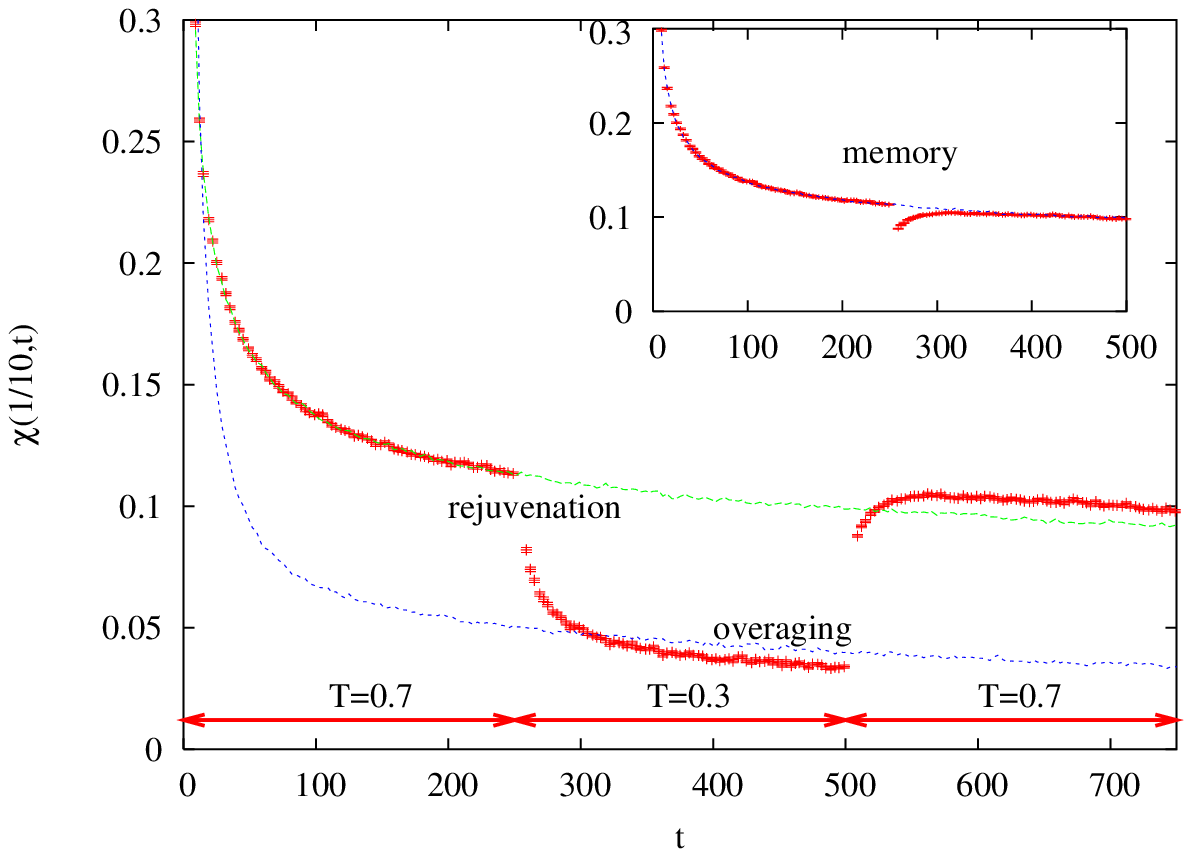}
\hspace{0.5cm}$^{({\rm b})}$\hspace{-1.1cm}
\includegraphics[width=0.53\textwidth]{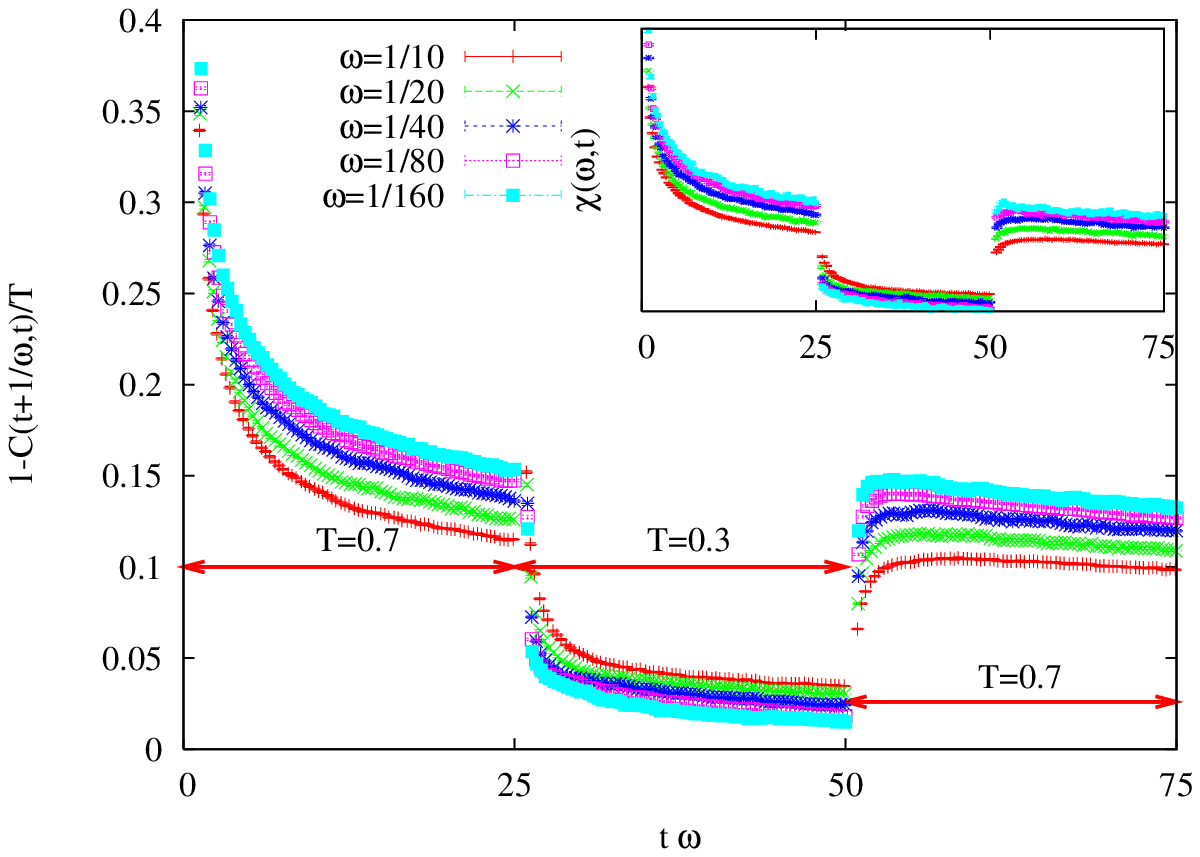}
\caption{\label{fig:2} Similar plots to Fig.\ref{fig:1} but for
  Kawasaki dynamics.  (a) Exact DC susceptibility in a $T$-cycle;
  memory and rejuvenation are both observed (in the inset we remove
  the part at $T_2$ to show better memory).  (b) As for Glauber, the
  rejuvenation signal vanishes at larger time scales, although in a
  much slower way.}
\end{figure}

This can be achieved easily just by switching to Kawasaki dynamics.  While it
is well known that domains then coarsen as $t^{1/3}$ at large
times~\cite{Bray_Review}, it has been shown recently that, due to initial
moves that requires thermal activation, the Kawasaki dynamics of the $2d$
ferromagnet get stuck for time scales shorter than $\tau =\exp(8
\beta)$~\cite{Kawasaki05} so that the grow is only logarithmic in this regime
(where the system is actually hardly distinguishable from a spin glass; even
its $T_{eff}$ resembles those of mean field disordered
systems~\cite{Kawasaki05}).  This strong temperature dependence of domains
growth is sufficient to add memory to our rejuvenation effect.  In
Fig.\ref{fig:2}(a), we now observe a quasi perfect memory effect due to the
freezing of the coarsening dynamics at $T_2$, so that back to $T_1$ the
dynamics continues where it has left (apart from a short, fast transient). We
also see overaging, another interesting effect observed in some experiments
(that we will not discuss here). The $2d$ Kawasaki model is probably the
simplest model in finite dimension that display these phenomena. This
demonstrates that memory and rejuvenation can be observed numerically even in
simple models without disorder. All that are good news, given the experimental
ubiquity of these effects but it again demonstrates that caution has to be
taken when interpreting such data. Indeed, the rejuvenation signal tends to
decrease, although slowly, at larger time scales (Fig.\ref{fig:2}(b)).

An alternative simple way to introduce such a $T$-dependence in the
dynamics is to add small disorder and/or frustration in the couplings
(or in the magnetic field), in which case the dynamics at $T_2$ could
again be slow enough to allow the observation of memory. The recipe
how to cook a model with memory and rejuvenation is thus quite
simple. This explains the results of~\cite{Jimenez}, where they
observed similar effects in site-diluted ferromagnet.  All these
results actually resemble what is experimentally observed in
disordered~\cite{ReEntrant} and frustrated magnets~\cite{NoDisorder},
probably because the underlying mechanism of interfaces pining is
similar.

\section{Temperature Cycle experiment in 2d XY model}

\begin{figure}
$^{({\rm a})}$\hspace{-1cm}
\includegraphics[width=0.53\textwidth]{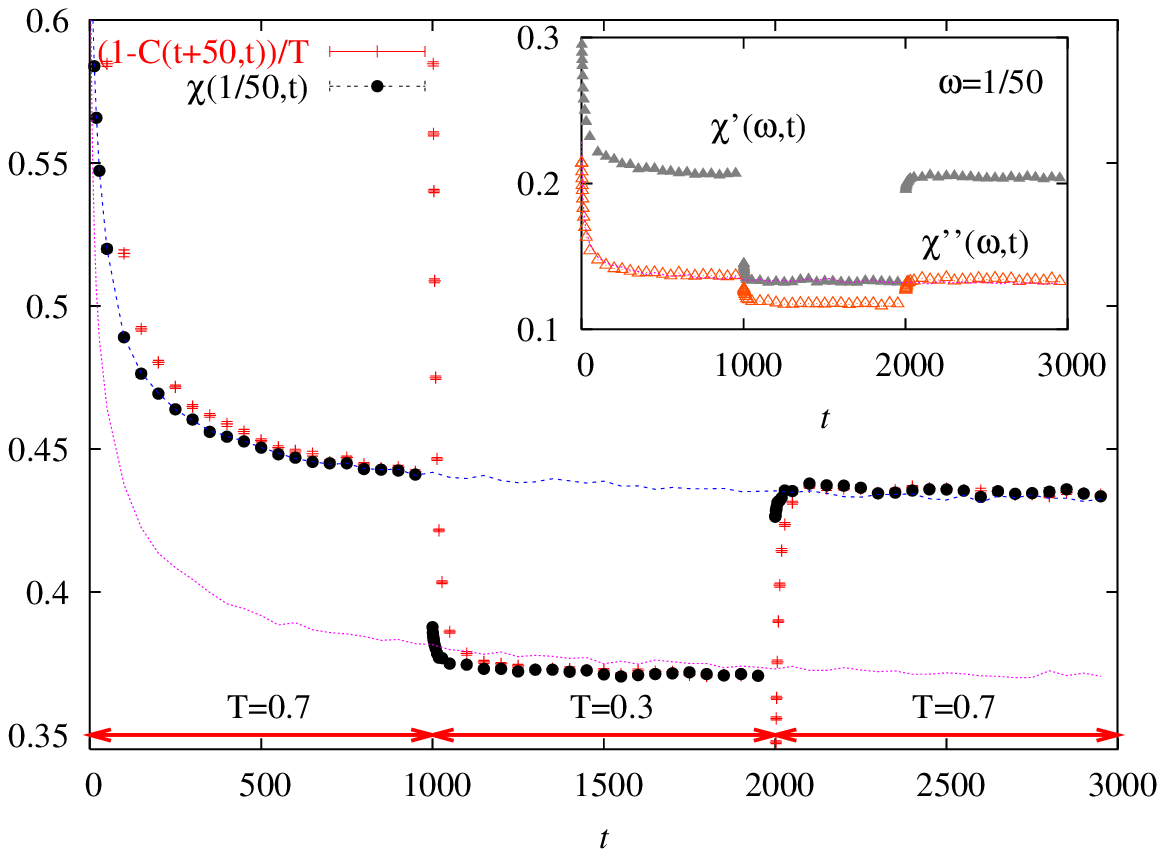}
\hspace{0.5cm}$^{({\rm b})}$\hspace{-1.1cm}
\includegraphics[width=0.53\textwidth]{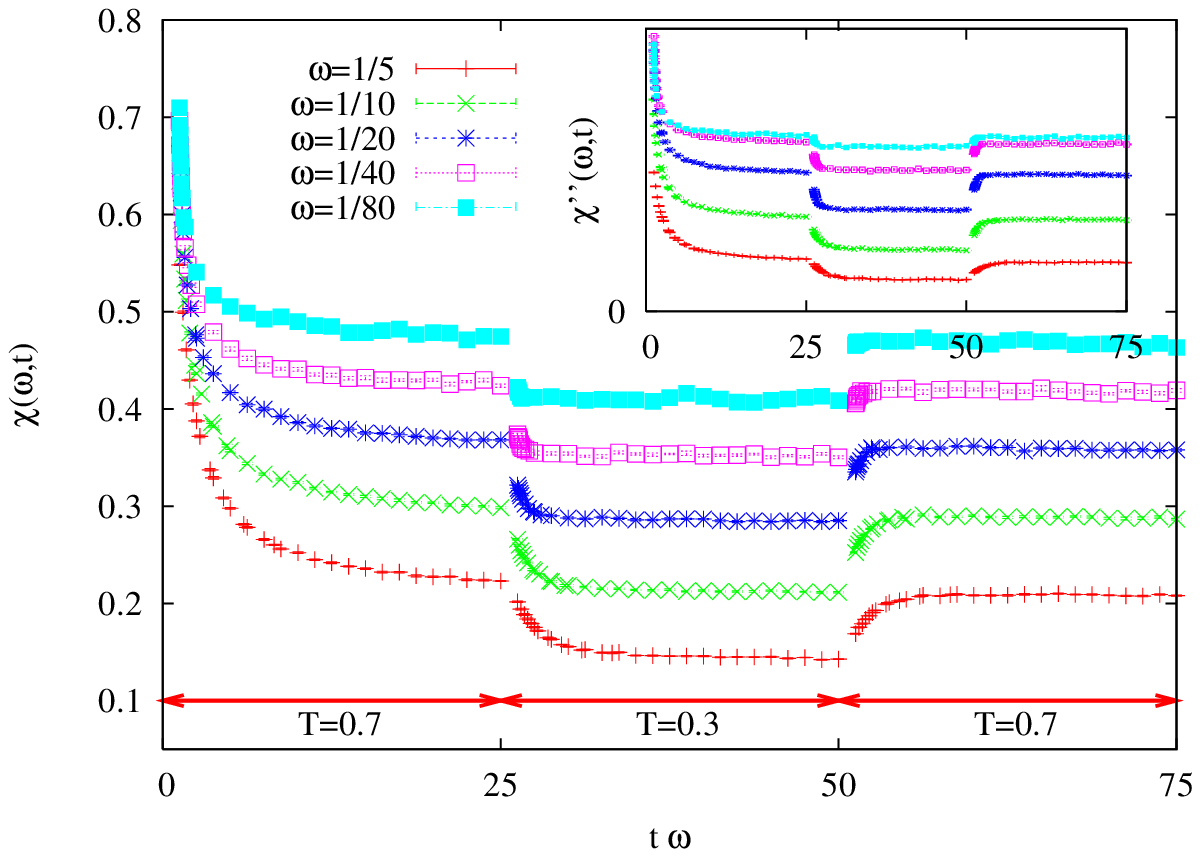}
\caption{\label{fig:3} $T$-cycling in the $2d$ XY model (same
  presentation of data as in Fig.\ref{fig:1}).\\ (a) The FDT-based
  susceptibility enhanced artificially the rejuvenation effect.\\ (b)
  Rejuvenation fastly vanishes at large time scales.}
\end{figure}

We turn briefly to the $2d$ XY model where the situation is quite
different: here we expect a rejuvenation signal from equilibrium
physics since the equilibrium correlation function essentially behaves
as $C(r) \propto r^{-T}$ so that all length scales have to be
re-equilibrated upon $T$-changes, as is evidenced by
Eq.(\ref{MyResult}).  This model thus seems to be a good illustration
of the ``many length scales'' ideas advocated in \cite{Review_cycle}.
It was suggested in~\cite{XY_Ludo} that the $2d$ XY model may capture
most of the experimental spin glass phenomenology, but our numerical
studies for this model are in disagreement with this
picture~\cite{FloFede}.  Firstly, as can be checked in the data of
Fig.\ref{fig:3}(a), the FDT violations we reported in the Ising model
are even stronger in the XY model, so that the impressive rejuvenation
effect previously seen in the correlation is actually very tiny for
the susceptibility. Secondly, due to the form of the correlation in
Eq.(\ref{MyResult}), which is typical of aging at criticality, the
large time limit makes all the rejuvenation effect to concentrate in a
vanishing small time window, see Fig.~\ref{fig:3}(b).

All this makes the result of~\cite{XY_Ludo} a bit artificial. While
the mechanism of re-adaptation at all scales is certainly relevant to
glassy dynamics, the use of the XY model is not really justified,
mainly because it is indeed a very special critical system, and
critical dynamics is quite different from the one observed in usual
aging. Unfortunately surfing on a critical line does not seem to be
sufficient to interpret spin glass experiments.

\section{Conclusion and discussion}

\begin{figure}
$^{({\rm a})}$\hspace{-1cm}
\includegraphics[width=0.53\textwidth]{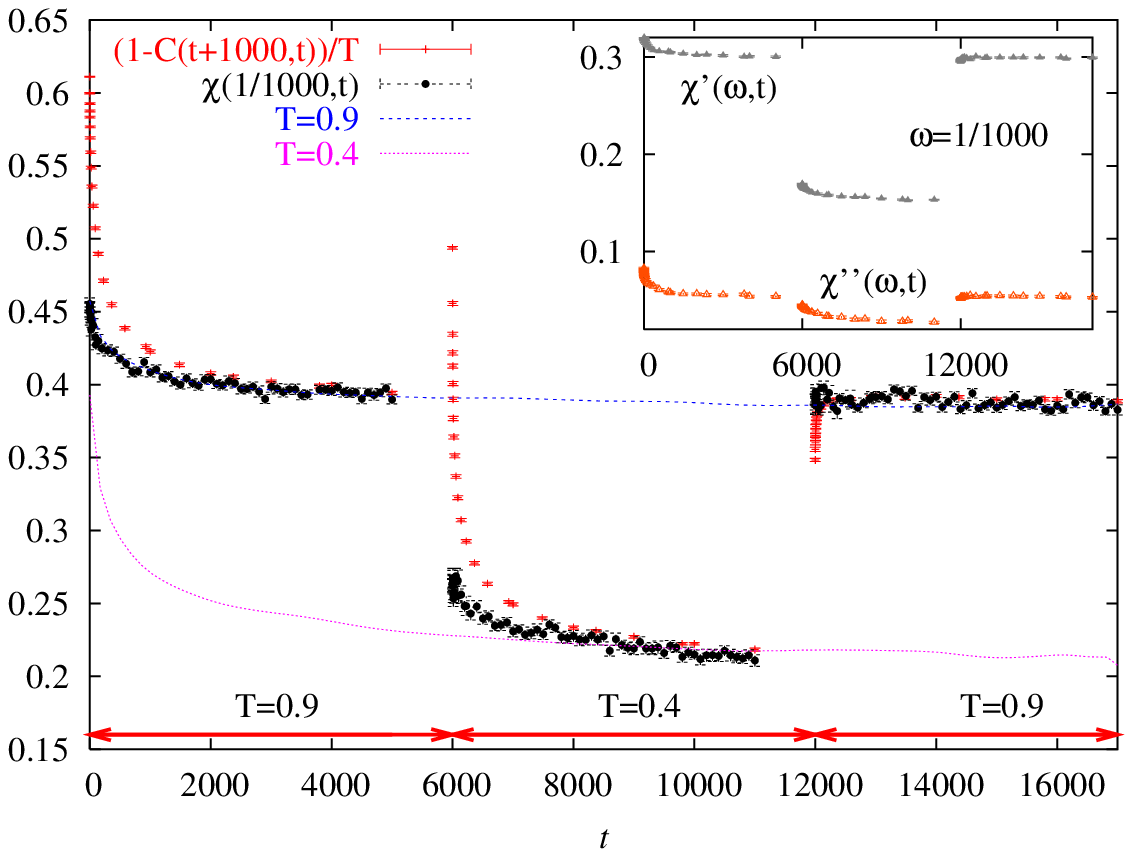}
\hspace{0.5cm}$^{({\rm b})}$\hspace{-1.1cm}
\includegraphics[width=0.53\textwidth]{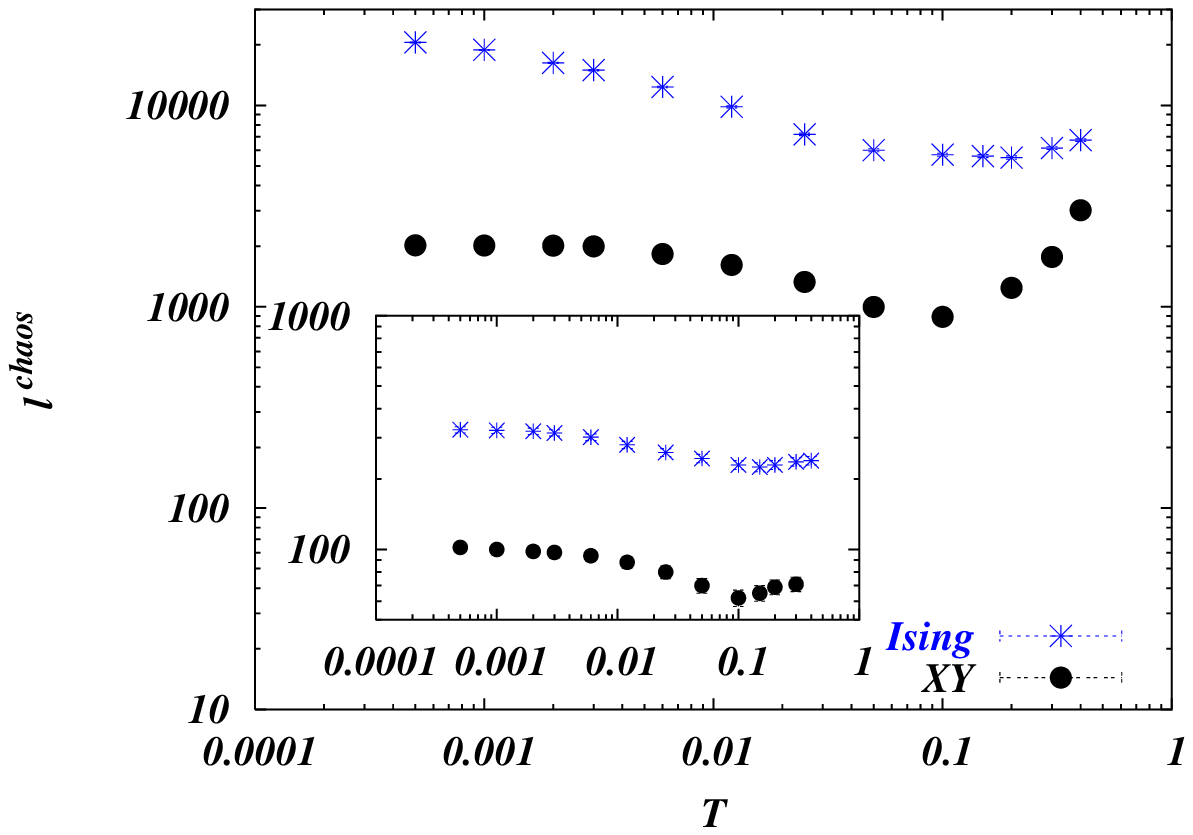}
\caption{\label{fig:4} (a) $T$-cycling in $4d$ spin glasses; data look
  very similar to what is obtained in the ferromagnets (in the inset
  data in oscillatory field). (b) Chaotic length, beyond which
  temperature chaos is observable, as a function of $T$ in the $3d$ XY
  and Ising spin glasses from real space renormalization (after
  \cite{ChaosXY04}, $\Delta T=0.001$ in the main plot, $\Delta T=0.01$
  in the inset). Chaos is much stronger for continuous spins.}
\end{figure}

Once again, studying simple models provided important lessons.
Firstly, contrary to what was believed, it is not so hard to observe
either rejuvenation and memory in simulations at finite times, in fact
even a simple ferromagnetic model can do that.  We also showed that
assuming FDT enhanced artificially the rejuvenation effect, and that
one can have a larger susceptibility at a lower temperature without
any restart of the dynamics.  Therefore careful interpretations of
simulation have to be made, and the long time limit has to be studied
before doing any comparison with experimental data. All these points
are valid for spin glasses, as can be checked in Fig.\ref{fig:4}(a).

In recent years, many authors concluded that since memory and rejuvenation can
be observed without temperature chaos, this concept is irrelevant (we saw
indeed that these effects can be obtained almost in any models if one tunes
properly the parameters).  Nevertheless the phenomenology we observed remains
quite far from what is observed in spin glass experiments when looking more
closely.  Firstly, the large time limit is different.  Secondly, in spin
glasses like AgMn the rejuvenation can be complete, so that the susceptibility
at $T_2$ after the stage at $T_1$ is the same as in a direct quench at $T_2$
(these are the only real rejuvenations according to \cite{Ghost}); it is
hardly the case for all the models considered here. It has been argued that
temperature changes are not instantaneous in experiments so that numerical
quenches have to be also progressive in order to obtain (maybe) a complete
rejuvenation~\cite{JP_Ludo,Jimenez}. We expect that this will not affect too
much the exact susceptibility, but rather its approximation based on FDT and
its ``unphysical'' part. It is finally important to mention the {\it second
  rejuvenation} effect that is sometime observed in Heisenberg spin glass when
heating back to $T_1$~\cite{shift,JapChaos,Ghost}, so that the dynamics looks
like quenched from a higher temperature in this reheating step.  This is not
observed in the models considered here, nor in simulations of spin glass, and
can neither be understood within the XY model, even qualitatively. This
suggests something new is at work, and this might well be {\em temperature
  chaos}. This is an old issue in spin glass community that for a while shared
many common points with the Loch Ness monster: many people talk about it, yet
no one really saw it.  It seems now that it exists in mean field as well as
finite dimensional systems~\cite{ChaosREREM,Chaos1,Chaos2} and it can also be
shown that the lengths beyond which chaotic effects are observable should be
shorter for continuous spins than for Ising spins~\cite{ChaosXY04}, see also
Fig.\ref{fig:4}(b), which would explain nicely why rejuvenation effects are
stronger for Heisenberg spin glasses, and why Ising samples do not seem to
display a second rejuvenation.

As usual, it would be certainly funny to look to these statements in a
few years, when most of these questions will have hopefully found
their answers.

\ack We would like to thank J.-P.~Bouchaud, V.~Dupuis, P.~J\"onnson,
E.~Vincent, O.~White, H.~Yoshino and L.~Zdeborov\'a for useful discussions
about these issues.

\end{document}